\documentclass{aa}  
\usepackage{graphicx}
\usepackage{natbib}

\bibpunct{(}{)}{;}{a}{}{,}   
\usepackage{txfonts}

\begin{document}
\title{Interferometric multi-wavelength (sub)millimeter continuum
  study of the young high-mass protocluster IRAS\,05358+3543}


   \author{H.~Beuther\inst{1},
          S. Leurini\inst{2,3},
          P. Schilke\inst{2},
          F. Wyrowski,
          K.M. Menten\inst{2}
          \and
          Q.~Zhang\inst{4}
          }

   \offprints{H.~Beuther}

   \institute{Max-Planck-Institute for Astronomy, K\"onigstuhl 17, 
              69117 Heidelberg, Germany\\
              \email{beuther@mpia.de}
         \and
              Max-Planck-Institute for Radioastronomy, Auf dem H\"ugel 69, 
              53121 Bonn, Germany\\
              \email{name@mpifr-bonn.mpg.de}
         \and              
              European Southern Observatory, Karl-Schwarzschild-Str. 2, 
              85748 Garching, Germany\\
              \email{sleurini@eso.org}
         \and
              Harvard-Smithsonian Center for Astrophysics, 60 Garden Street,
              Cambridge, MA 02138, USA\\
             \email{qzhang@cfa.harvard.edu}
             }
\authorrunning{Beuther et al.}
\titlerunning{(Sub)mm observations of IRAS\\05358+3543}

   \date{}

  \abstract
    {}
    {We study the small-scale structure of massive
      star-forming regions through interferometric observations in several
      (sub)mm wavelength bands. These observations resolve multiple sources,
      yield mass and column density estimates, and give information
      about the density profiles as well as the dust and temperature
      properties.}
    {We observed the young massive star-forming region
      IRAS\,05358+3543 at high-spatial resolution in the continuum
      emission at 3.1 and 1.2\,mm with the Plateau de Bure
      Interferometer, and at 875 and 438\,$\mu$m with the
      Submillimeter Array. The observations are accompanied by VLA
      3.6\,cm archival continuum data.}
    {We resolve at least four continuum sub-sources that are likely of
      protostellar nature. Two of them are potentially part of a
      proto-binary system with a projected separation of 1700\,AU.
      Additional (sub)mm continuum peaks are not necessarily harboring
      protostars but maybe caused by the multiple molecular outflows.
      The spectral energy distributions (SEDs) of the sub-sources show
      several features. The main power house mm1, which is associated
      with CH$_3$OH maser emission, a hypercompact H{\sc ii} region
      and a mid-infrared source, exhibits a typical SED with a
      free-free emission component at cm and long mm wavelengths and a
      cold dust component in the (sub)mm part of the spectrum
      (spectral index between 1.2\,mm and 438\,$\mu$m $\alpha\sim
      3.6$). The free-free emission corresponds to a Lyman continuum
      flux of an embedded 13\,M$_{\odot}$ B1 star.  The coldest source
      of the region, mm3, has $\alpha\sim 3.7$ between 1.2\,mm and
      875\,$\mu$m, but has lower than expected fluxes in the shorter
      wavelength 438\,$\mu$m band. This turnover of the
      Planck-function sets an upper limit on the dust temperature of
      mm3 of approximately 20\,K. The uv-data analysis of the density
      structure of individual sub-cores reveals distributions with
      power-law indices between 1.5 and 2. This resembles the density
      distributions of the larger-scale cluster-forming clump as well
      as those from typical low-mass cores. {{\it The files for
          figures 2, 3, 4 \& 6 are also available in electronic form
          at the CDS via anonymous ftp to cdsarc.u-strasbg.fr
          (130.79.128.5) or via
          http://cdsweb.u-strasbg.fr/cgi-bin/qcat?J/A+A/.}}}
   {}
   

   \maketitle
%

\section{Introduction}
\label{intro}

Single-dish (sub)mm continuum imaging in high-mass star-forming
regions revealed various interesting properties of the dust and gas
clumps harboring the star-forming clusters. The derived physical
parameters include the gas masses, the gas column densities, the
spectral indices and resulting dust properties as well as the density
distributions of the gas on cluster scales
\citep{hatchell2000,vandertak2000,molinari2000,mueller2002,beuther2002a,hatchell2003,walsh2003,williams2004,fontani2005,williams2005,hill2005,klein2005,faundez2004,beltran2006}.
However, all these studies had insufficient spatial resolution (at
best 16000\,AU at 2\,kpc distance) to resolve sub-structure within the
star-forming regions, and thus were inadequate to study the properties
of individual cluster members or unresolved multiple systems. A
further limitation was that many of such studies only covered one
frequency band (870\,$\mu$m or 1.2\,mm) and hence could not
investigate the spectral index of the gas emission, which allows to
set constraints on the dust properties.

To overcome some of these limitations we conducted a comprehensive
program observing the very young massive star-forming region
IRAS\,05358+3543 at high spatial resolution with the Submillimeter
Array (SMA\footnote{The Submillimeter Array is a joint project between
  the Smithsonian Astrophysical Observatory and the Academia Sinica
  Institute of Astronomy and Astrophysics, and is funded by the
  Smithsonian Institution and the Academia Sinica.}) in the 438 and
875\,$\mu$m band, and with the IRAM Plateau de Bure Interferometer
(PdBI\footnote{IRAM is supported by INSU/CNRS (France), MPG (Germany),
  and IGN(Spain).})  in the 1.2 and 3.1\,mm bands.  These observations
resolve the high-mass star-forming cluster into various sub-components
at four different wavelengths, allowing to study the physical
properties of individual cluster members or still unresolved, likely
gravitationally bound multiple systems in great detail.

\begin{figure}[htb]
\includegraphics[angle=-90,width=8.8cm]{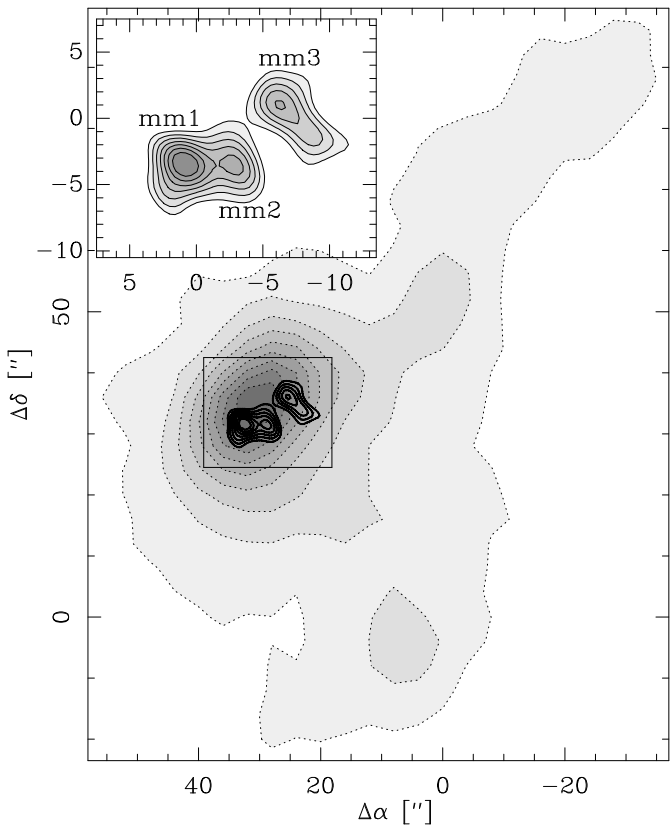}
\caption{Overview of the region.  The large emission shown in
  grey-scale with dotted contours is the 1.2\,mm continuum emission
  observed with the MAMBO array at the IRAM\,30m telescope
  \citep{beuther2002a}. The full contours (enlarged in the inlay-box)
  show the 2.6\,mm continuum data from the PdBI at a spatial
  resolution of $4.0''\times 3.0''$ with the labeled three identified
  mm continuum sources \citep{beuther2002d}.  The 1.2\,mm MAMBO data
  are contoured from 106\,mJy\,beam$^{-1}$ to 954\,mJy\,beam$^{-1}$
  (step 106\,mJy\,beam$^{-1}$), the 2.6\,mm PdBI are contoured from
  6.6\,mJy\,beam$^{-1}$ to 22.0\,mJy\,beam$^{-1}$ (step
  2.2\,mJy\,beam$^{-1}$). The (0/0) position of the large map is the
  position of the IRAS\,05358+3543 source which is derived from the
  IRAS 12\,$\mu$m image, whereas the mm continuum peak correlates with
  the IRAS 100\,$\mu$m peak. The (0/0) position of the inlay is the
  phase center of the older observations by \citet{beuther2002d}.}
\label{05358_intro}
\end{figure}

IRAS\,05358+3543 is part of a sample of 69 High-Mass Protostellar
Objects (HMPOs) investigated in depth over the last few years by
\citet{sridha,beuther2002a,beuther2002b,beuther2002c,williams2004,williams2005,fuller2005}.
At a kinematic distance of 1.8\,kpc, the bolometric luminosity of the
region is $10^{3.8}$\,L$_{\odot}$, and the integrated gas mass is
$\sim$300\,M$_{\odot}$\footnote{The cited gas masses are at a 50\%
  level of the masses originally presented in
  \citet{beuther2002a,beuther2002d} because of different assumptions
  for the dust opacities \citep{beuther2002erratum}.}.  H$_2$O and
Class {\sc ii} CH$_3$OH maser emission was studied at high spatial
resolution \citep{tofani1995,minier2000,beuther2002c}, but so far no
cm continuum emission had been detected down to the threshold of
1\,mJy, indicative of the the very young evolutionary stage of the
region.  Furthermore, high-spatial-resolution interferometric
observations of the region revealed at least three molecular outflows,
one of them being the most collimated massive molecular outflow known
today \citep{beuther2002d}.  These observations also resolved the
evolving protocluster at 2.6\,mm wavelength with a $4''\times 3''$
beam into three sub-sources (mm1 to mm3) with gas masses between 37
and 50\,M$_{\odot}\,^3$.  Fig.~\ref{05358_intro} gives an overview of
the large and small-scale dust continuum emission from previous
observations \citep{beuther2002a,beuther2002d}. Two of the three
identified outflows emanate from the vicinity of mm1. McCaughrean et
al.~(in prep.)  detected with Keck observations two mid-infrared
sources in the region, one being spatially coincident with the main mm
continuum peak mm1. This mid-infrared source was recently resolved by
\citet{longmore2006} into a double-source separated by $1''$
corresponding to 1800\,AU. Near-infrared polarimetric imaging also
revealed a deeply embedded source associated with mm1 \citep{yao2000}.
Previously identified infrared sources, some of them were suggested to
be the outflow driving sources \citep{porras2000}, were shown to be
offset from the main mm and mid-infrared sources and hence unlikely to
be the main power houses in the region \citep{beuther2002d}.

In this paper, we present the (sub)mm continuum observations from all
four wavelengths bands. The corresponding spectral line data will be
published separately by Leurini et al.~(in prep.). For the
  nomenclature of this paper, we use the term protostar (and
  corresponding proto-cluster, proto-binary) for massive young stellar
  objects that are still actively accreting, independent of whether they
  have started hydrogen burning already or not.

\section{Observations and Data Reduction}
\label{obs}

We observed IRAS\,05358+3543 with three interferometers (SMA, PdBI and
VLA) in ten different configurations or spectral setups. The most
important observational parameters are given in Table
\ref{observationparameters}. The velocity of rest was
$v_{\rm{lsr}}=-17.6$\,km\,s$^{-1}$, and the flux accuracy of the
observations is estimated to be good within 20\%. Below we discuss a
few additional details to the different observations.

\subsection{Submillimeter Array Observations at 875\,$\mu$m}

The SMA observations in the compact and extended configuration were
centered at 348/338\,GHz (upper and lower sideband, USB/LSB,
875\,$\mu$m). The projected baselines ranged between 15 and
252\,k$\lambda$. The short baseline cutoff implies that source
structures $\geq 16''$ are filtered out by the observations. The phase
center of the observations was R.A.(J2000.0) 05$^h$39$^m$13$^s$.07 and
decl.(J2000.0) $+$35$^{\circ}$45$'$51$''$.2. The zenith opacities,
measured with the NRAO tipping radiometer located at the Caltech
Submillimeter Observatory, were good during both tracks with
$\tau(\rm{348GHz})\sim 0.18$ (scaled from the 225\,GHz measurement via
$\tau(\rm{348GHz})\sim 2.8\times \tau(\rm{225GHz})$). The correlator
had a bandwidth of 1.968\,GHz and the channel separation was 0.8125\,MHz.
Measured double-sideband system temperatures corrected to the top of
the atmosphere were between 150 and 500\,K, mainly depending on the
elevation of the source. The initial flagging and calibration was done
with the IDL superset MIR originally developed for the Owens Valley
Radio Observatory \citep{scoville1993} and adapted for the
SMA\footnote{The MIR cookbook by Charlie Qi can be found at
  http://cfa-www.harvard.edu/$\sim$cqi/mircook.html.}. The imaging and
data analysis was conducted in MIRIAD \citep{sault1995}. During the
observations on November 11th, 2004, the position of the primary
calibrator 3C111 was wrong in the catalog by $\sim$0.6$''$.
Therefore, we self-calibrated our secondary phase calibrator 0552+398
($16.3^{\circ}$ from the source) shifting it in the map to the correct
position. The solutions were then applied to IRAS\,05358+3543. The
875\,$\mu$m continuum data were produced by averaging the apparently
line-free part of the USB and LSB data. The corresponding line
emission will be discussed in a separate paper by Leurini et al.~(in
prep.).

\subsection{Submillimeter Array Observations at 438\,$\mu$m}

The weather conditions for the 438$\mu$m/690GHz observations were
excellent with a zenith opacity $\tau(\rm{230GHz})$ between 0.03 and
0.05 throughout the night. This corresponds to zenith opacities at
690\,GHz between 0.6 and 1.0 ($\tau(\rm{690GHz})\sim 20\times
(\tau(\rm{230GHz})-0.01)$, \citealt{masson1994}). The phase center and
correlator setup were the same as for the 875\,$\mu$m observations.
The primary beam of the SMA at these high frequencies is $\sim 18''$.
The covered frequency ranges were 679.78 to 681.75\,GHz and 689.78 to
691.75\,GHz. Measured double-sideband system temperatures corrected to
the top of the atmosphere were between 1350 and 10000\,K, depending on
the elevation of the source and the opacity. The sensitivity was
limited by the strong side-lobes of the strongest emission peaks.
Since the third main mm continuum source is approximately at the edge
of the SMA primary beam at this frequency, we applied a correction for
the primary beam attenuation to the final image to determine the
fluxes more properly. Although the synthesized beam of the continuum
image is $1.4''\times 0.9''$, we work only on a smoothed version with
a $1.9''\times 1.4''$ beam, comparable to the 3\,mm data.

\subsection{Plateau de Bure Interferometer Observations at 1.2 and 3.1\,mm}
\label{pdbi_obs}

The PdBI was used in two different frequency setups in 2003 and 2005.
A first frequency setup was performed in four tracks between January
and October 2003 in the B-C and D configurations of the array.  The
reference center is R.A.~(J2000) 05$^h$39$^m$13$^s$.0 and
decl.~(J2000) 35$^{\circ}$45$'$54$''$.  The 3\,mm receivers were used
in single sideband mode and tuned to 96.6~GHz; the 1\,mm receivers, in
double sideband mode, were tuned to 241.85~GHz, (USB). One 320~MHz
unit was placed to obtain a continuum measurement at 3.1\,mm. Four
overlapping 160\,MHz units were used to obtain the 1.2\,mm continuum
measurement.  The system temperatures varied between 100 and 150\,K in
the 3.1\,mm band, and between 300 and 600\,K in the 1.2\,mm band. The
second frequency setup was observed in 2005 in the A and B
configurations.  The 1mm receivers were tuned in lower sideband to
241.2\,GHz.  Three 160\,MHz units were used to get the continuum
measurement.  Measured system temperatures in the 1.2\,mm receivers
ranged between 240 and 1000\,K, with one receiver measuring system
temperatures of 900\,K. The data calibration were performed in CLIC,
and the imaging in MAPPING, both packages are part of the GILDAS
software suite\footnote{See http://www.iram.fr/IRAMFR/GILDAS}.

\subsection{Very Large Array archival data at 3.6\,cm}
\label{vla}

Searching the Archive of the Very Large Array (VLA), we found A+B
array data observed at 3.6\,cm wavelength in 2002 and 2003 (project
AR482). The data were reduced with the standard AIPS procedures.

\begin{table*}[htb]
\caption{Main observation parameters}
\begin{center}
\begin{tabular}{lrrrrrrrr}
\hline \hline
$\lambda$ & Obs & Date & Config.& \multicolumn{3}{c}{Calibrators} & $\Theta$ & $1\sigma$ \\
          &     &      &       & Phase/Amp.   & Flux dens. & Bandpass \\
(mm) & & & & & & & $('')$ & $(\frac{\rm{mJy}}{\rm{beam}})$ \\
\hline
36  & VLA  & 08/02-06/03 & AB  & 0555+398 & 3C286 & 0555+398 & $0.49\times 0.41$ & 0.018 \\
3.1 & PdBI & 01/03-10/03 & BCD & 0528+134 & NRAO150 & 0420-014 & $1.9\times 1.4''$ & 0.3 \\
    &      &             &     & 552+398 & 0528+134 & 3C454.2\\
    &      &             &     &         &          & NRAO150\\
1.2 & PdBI & 01/03-10/03 & BCD & 0528+134 & NRAO150 & 0420-014  & $1.26\times 0.84$ & 2.3 \\
    &      &             &     & 0552+398 & 0528+134& 3C454.2\\
    &      &             &     &         &          & NRAO150\\
1.2 & PdBI & 02/05       & AB  & 0529+483 & 3C273 & 3C84 & $0.6\times 0.44$ & 1.4\\
    &      &             &     & J0418+380 & 3C273\\
0.875 & SMA & 11/04 & comp. & 3C111 & Callisto & Jupiter & $1.14\times 0.57$ & 7 \\
       &      &          &         &       &          & \& Callisto \\
0.875 & SMA & 01/05 & ext. & 3C111 & Callisto & Jupiter & $1.14\times 0.57$ & 7\\
      &     &          &      &       & 3C279 & 3C279 \\ 

0.438 & SMA & 01/06 & comp. & Vesta & Callisto & Callisto &$1.4\times 0.9$ &  400 \\
      &     &          &       &       &          & \& Uranus\\
\hline \hline
\end{tabular}
\end{center}
\footnotesize{The Table shows the observing frequency $\lambda$, the observatories, the dates of observation, the array configurations, the calibrators, the resulting synthesized beam $\Theta$ and the $1\sigma$ value of the final maps.}
\label{observationparameters}
\end{table*}

\section{Results}

\subsection{Small-scale dust/gas distribution}
\label{small-scale}

\begin{figure*}[htb]
\begin{center}
\includegraphics[angle=-90,width=13cm]{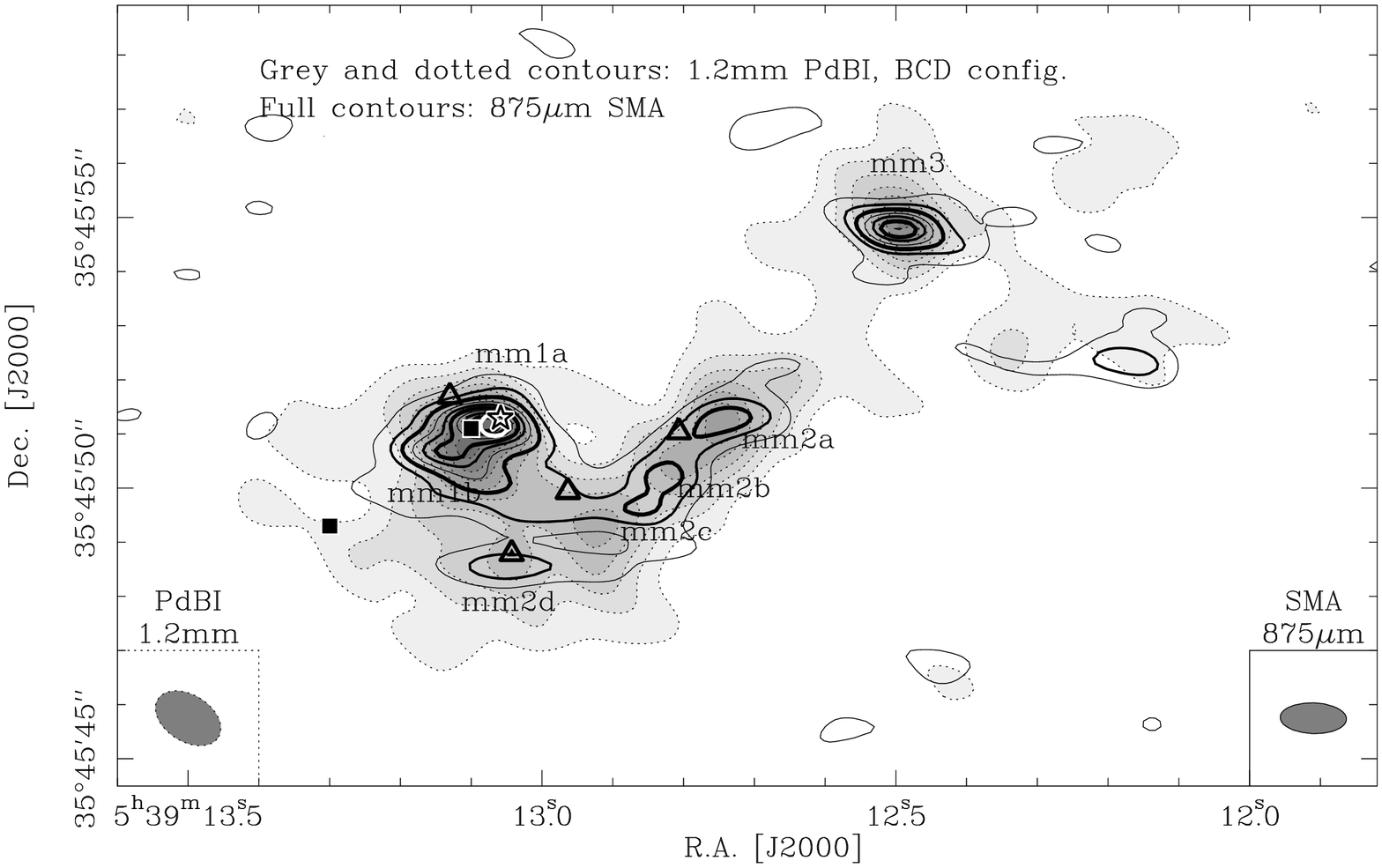}
\includegraphics[angle=-90,width=13cm]{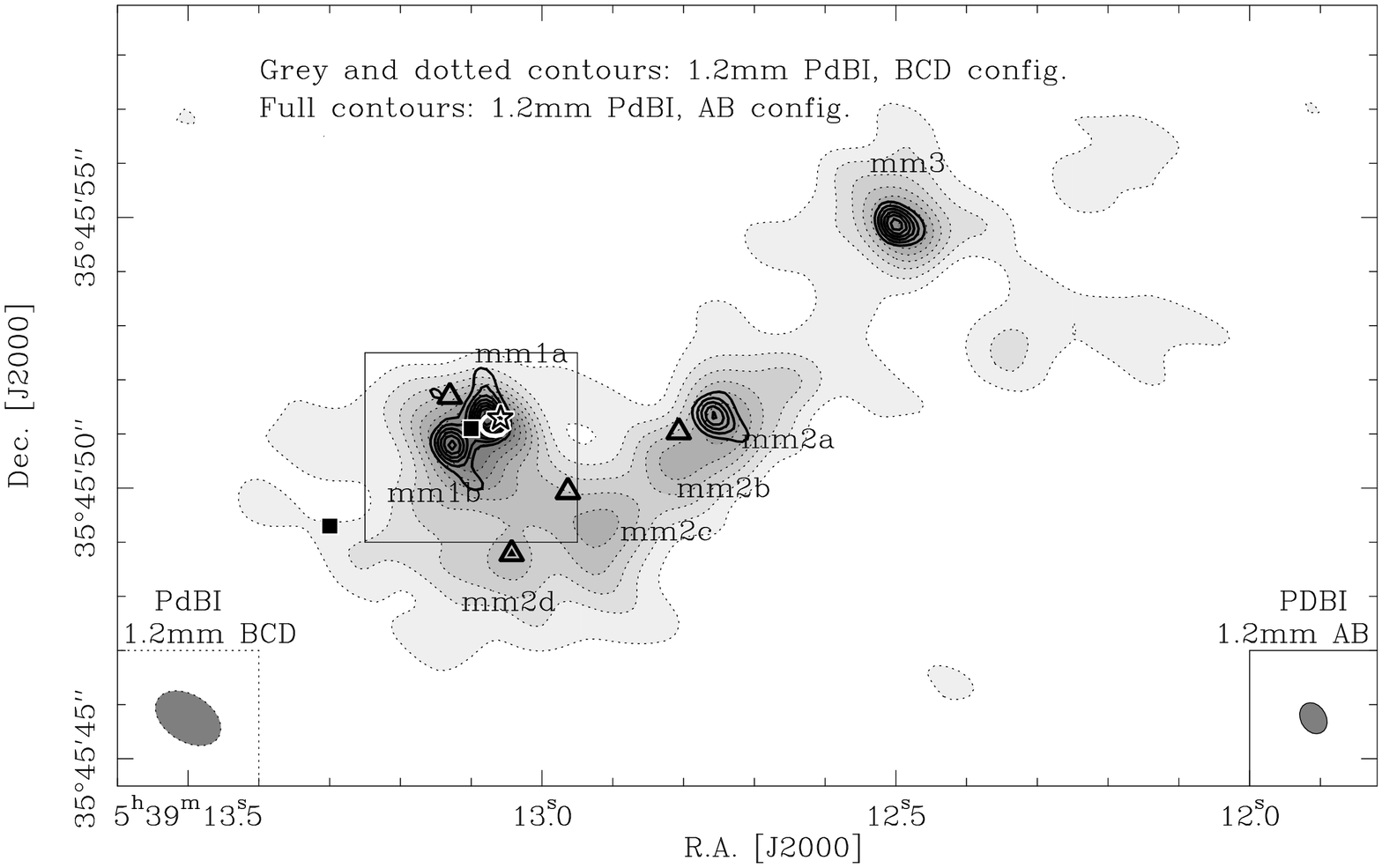}
\end{center}
\caption{(Sub)mm continuum images toward IRAS\,05358+3543 at 1.2\,mm
  (PdBI) and at 875\,$\mu$m (SMA). The top panel shows in grey-scale
  with dotted contours the 1.2\,mm data (BCD configuration) and in
  full contours with increasing thickness the 875\,$\mu$m data. The
  contouring is done from 10 to 90\% of the peak emission of each
  image (step 10\%), with peak flux values of 58.9 and
  164.3\,mJy\,beam$^{-1}$ for the 1.2\,mm and 875\,$\mu$m bands,
  respectively. The bottom-panel shows in grey-scale with dotted
  contours again the same 1.2\,mm data (BCD configuration), and in
  full contours the 1.2\,mm data observed at even higher spatial
  resolution ($0.6''\times 0.4''$) in the AB configuration. The
  contouring of the latter dataset is from 20 to 90\% (step 10\%) of
  the peak flux 29.8\,mJy\,beam$^{-1}$. In both panels, the white
  ellipse shows the VLA 3.6\,cm 50\% contour level (0.31\,mJy).  The
  star, triangles and squares mark the positions of the Class {\sc ii}
  CH$_3$OH maser \citep{minier2000}, the H$_2$O masers (open triangles
  from \citealt{tofani1995} and small filled triangle from
  \citealt{beuther2002c}) and the mid-infrared sources (McCaughrean et
  al.~in prep.). The newly identified double-source mm1a and mm1b is
  labeled as well. The box outlines the region shown in Figure
  \ref{zoom}. {\it The files for this figure are also available in
    electronic form at the CDS via anonymous ftp to
    cdsarc.u-strasbg.fr (130.79.128.5) or via
    http://cdsweb.u-strasbg.fr/cgi-bin/qcat?J/A+A/.}}
\label{1.3mm_850mu}
\end{figure*}

The high-spatial-resolution multi-frequency data allow to disentangle
considerable sub-structure in this massive young star-forming cluster.
Fig.~\ref{1.3mm_850mu} presents overlays of the 1.2\,mm and the
875\,$\mu$m dust continuum emission. The top-panel shows images at
both wavelengths with comparable resolution (BCD configuration with the
PdBI at 1.2\,mm, and extended+compact configuration with the SMA at
875\,$\mu$m). The 3.1\,mm and 438\,$\mu$m images with slightly lower
spatial resolution will be discussed below (\S \ref{multi}). We
clearly resolve many sub-features compared to the more simple three
sources previously detected at lower spatial resolution (see
Fig.~\ref{05358_intro} and \citealt{beuther2002d}).  Furthermore, we
have additional 1.2\,mm continuum data with even higher spatial
resolution ($0.6''\times 0.4''$, AB configuration with the PdBI) shown
in the bottom-panel of Fig.~\ref{1.3mm_850mu}. While this
highest-spatial-resolution dataset filters out nearly all large-scale
emission it clearly identifies the most compact sources. We could have
combined all four configurations, but since the most extended
configurations dominate this image, we refrained from that.

\begin{figure}[htb]
\begin{center}
\includegraphics[angle=-90,width=8cm]{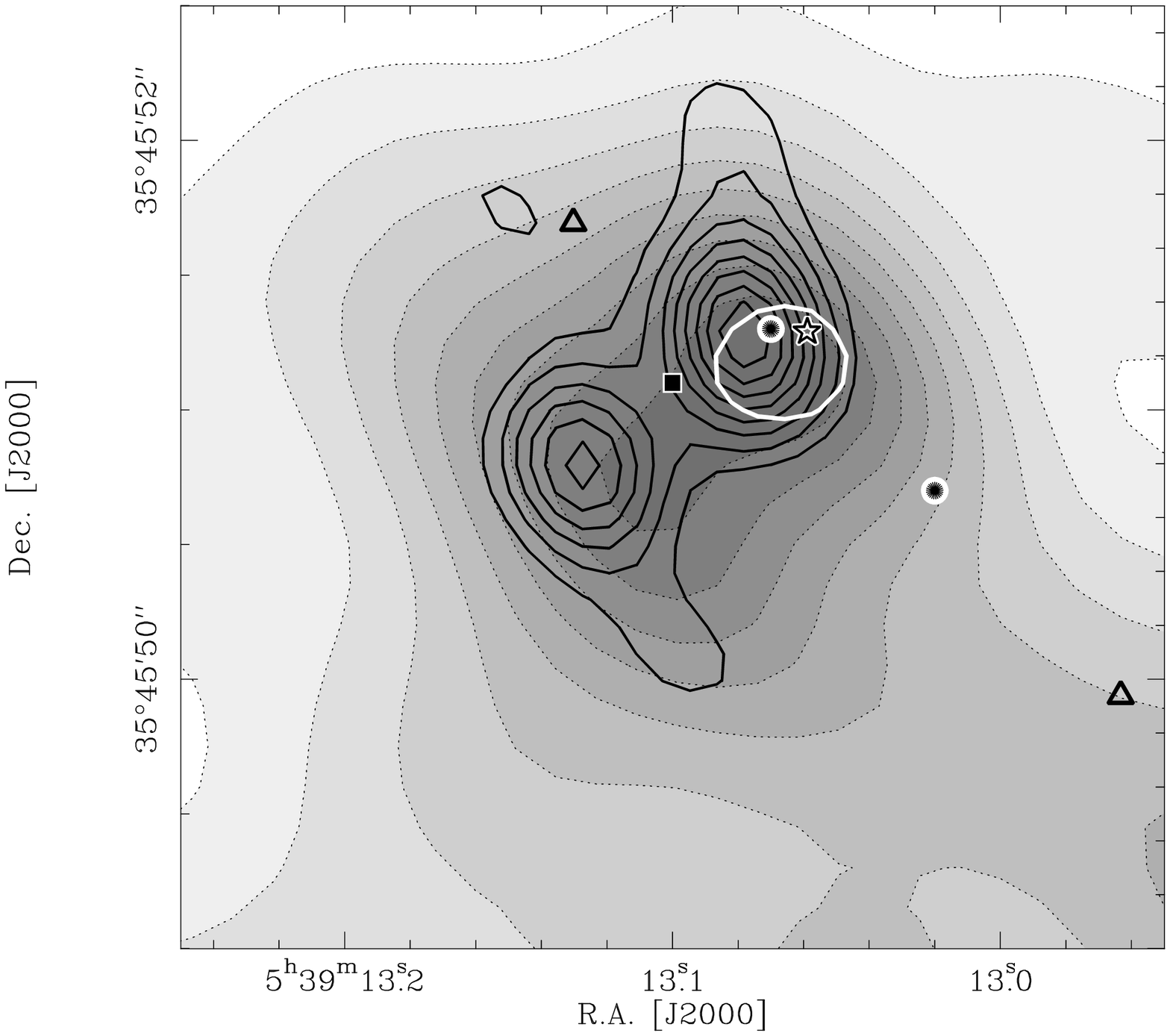}
\end{center}
\caption{Zoom into the central region of mm1a and mm1b as shown in
  Figure \ref{1.3mm_850mu}. The grey-scale with dotted contours shows
  again the same 1.2\,mm data (BCD configuration), and in full
  contours the 1.2\,mm data observed at higher spatial resolution
  ($0.6''\times 0.4''$) in the AB configuration. The contouring is the
  same as in Figure \ref{1.3mm_850mu}. The white ellipse shows the VLA
  3.6\,cm 50\% contour level (0.31\,mJy).  The star, triangles and
  squares mark the positions of the Class {\sc ii} CH$_3$OH maser
  \citep{minier2000}, the H$_2$O masers \citep{tofani1995} and the
  mid-infrared sources (McCaughrean et al.~in prep.). The
  black-and-white circles show the double mid-infrared sources
  discussed by \citet{longmore2006}. The $0.5''$ positional difference
  between the main mid-infrared peak positions by McCaughrean et al.
  and Longmore et al. can be accounted for by relatively poor pointing
  accuracies and the difficulties of registering mid-infrared data.
  {\it The files for this figure are also available in electronic form
    at the CDS via anonymous ftp to cdsarc.u-strasbg.fr (130.79.128.5)
    or via http://cdsweb.u-strasbg.fr/cgi-bin/qcat?J/A+A/.}}
\label{zoom}
\end{figure}

The main (sub)mm continuum source mm1, which is at the center of two
molecular outflows \citep{beuther2002d}, is resolved into two separate
(sub)mm continuum peaks, mm1a and mm1b. This is already observable at
$\sim$1$''$ in Fig.~\ref{1.3mm_850mu} (top panel) but more clearly
discernable in the highest-resolution data in the bottom panel of
Fig.~\ref{1.3mm_850mu} and the zoom in Fig.~\ref{zoom}. The compact
3.6\,cm continuum source and the Class {\sc ii} CH$_3$OH maser are
associated with mm1a. The cm continuum source is unresolved at the
given spatial resolution with a peak flux of 0.62\,mJy\,beam$^{-1}$
(below the detection limit of 1\,mJy by \citealt{sridha}), and
we only show the 50\% contour level in Figs.~\ref{1.3mm_850mu} \&
\ref{zoom}.  The closest H$_2$O maser feature detected by
\citet{tofani1995} is $\sim$1$''$ north-east of mm1a and not clearly
associated with one or the other peak as well. It should be noted that
this feature was not detected by \citet{beuther2002c}, probably due to
variability of the feature.  The projected separation between mm1a and
mm1b amounts to $\sim$0.94$''$, corresponding to a projected linear
separation of $\sim$1700\,AU at the given distance (1.8\,kpc).  The
main mid-infrared source detected by \citet{longmore2006} coincides
well with the position of mm1a (Fig.~\ref{zoom}). However, while the
secondary source mm1b is $\sim$1700\,AU toward the south-east, the
secondary mid-infrared source is $\sim$1800\,AU south-west of mm1a.
Hence, mm1b and the secondary mid-infrared source are apparently
independent indicating even more sub-structure in the evolving young
cluster.

The previously identified source mm2 is resolved even more, at least
into two sub-sources mm2a and mm2b (Fig.~\ref{1.3mm_850mu}). However,
while mm2a shows the same peak position in the 1.2\,mm band and the
875\,$\mu$m band, the peaks associated with mm2b are separated by
approximately $0.4''$. The mm peak mm2c is identified in the 1.2\,mm
image but not at shorter wavelengths. Inspecting the longer wavelength
3.1\,mm data in Figure \ref{sample}, one finds an elongation in this
lower-spatial-resolution dataset toward the mm2c position as well
(labeled ``jet?'' in Fig.~\ref{sample}). Since also the spectral index
between the 1.2\,mm and 875\,$\mu$m toward this position is very low
(of the order 1.5, see \S \ref{spectral_distribution} and
corresponding Fig.~\ref{ratiomap}) this feature may be due to an
underlying outflow/jet. The feature mm2d in Fig.~\ref{1.3mm_850mu} is
again discernable in both bands. One of the H$_2$O maser features by
\citet{tofani1995} is associated with this latter (sub)mm continuum
emission feature.

The third source mm3 remains a single compact object even at the
highest spatial resolution. The lower-resolution 3.1\,mm image in
Fig.~\ref{05358_intro} (inlay) shows an elongation from mm3 toward the
south-west, and the higher-resolution images in Fig.~\ref{1.3mm_850mu}
show weak emission there as well. Since these weak emission features
show different peak positions in the 1.2\,mm and the 875\,$\mu$m
bands, we refrain from further discussion of this feature.

Protostellar condensations should be detectable in all (sub)mm
wavelengths bands and not be filtered out in very extended
configurations. Therefore, we identify at $\sim$0.5$''$ spatial
resolution ($\sim 900$\,AU) four compact protostellar sources in the
region (mm1a, mm1b, mm2a, mm3) , the positions are listed in Table
\ref{positions}.

\begin{table}[htb]
\caption{Positions of the four identified protostellar sources}
\begin{center}
\begin{tabular}{lrr}
\hline
\hline
Source & R.A.~[J2000] & Dec.~[J2000] \\
\hline
mm1a & 05:39:13.08 & 35:45:51.3 \\
mm1b & 05:39:13.13 & 35:45:50.8 \\
mm2a & 05:39:12.76 & 35:45:51.3 \\
mm3  & 05:39:12.50 & 35:45:54.9 \\
\hline
\hline
\end{tabular}
\end{center}
\label{positions}
\end{table}

\subsection{The multi wavelength dataset}
\label{multi}

\begin{figure}[htb]
\begin{center}
\includegraphics[angle=-90,width=7cm]{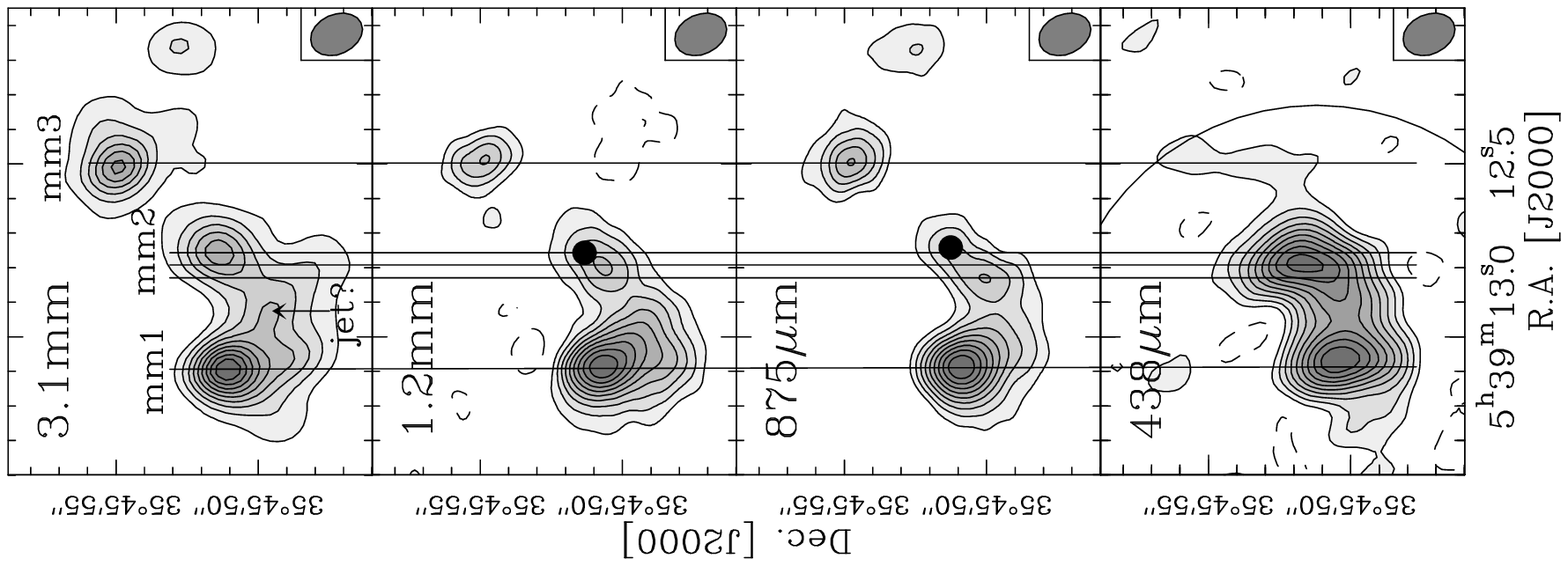}
\end{center}
\caption{(Sub)mm continuum images in the four observed bands smoothed
  to the spatial resolution of $1.9''\times 1.4''$ of the lowest resolution
  image at 3.1\,mm. The contour levels are from 10 to 90\% (step 10\%) of the
  peak emissions with peak fluxes of 10.3, 91.8, 270.8, and
  3569\,mJy\,beam$^{-1}$ for the 3.1\,mm, 1.2\,mm, 875, and 438\,$\mu$m bands,
  respectively. The 438\,$\mu$m image shown here is not primary-beam corrected
  yet, the size of the primary beam is indicated by the large half-circle in
  the bottom panel. The left and right vertical lines mark the positions of
  mm1 and mm3. In contrast, the peak position of mm2 shifts with wavelengths
  in these lower-resolution images which is indicated by the three middle
  vertical lines. However, the full circles mark the mm2a peak position in the
  corresponding high-resolution images (Fig.~\ref{1.3mm_850mu}). {\it The files for this figure are also available in
    electronic form at the CDS via anonymous ftp to
    cdsarc.u-strasbg.fr (130.79.128.5) or via
    http://cdsweb.u-strasbg.fr/cgi-bin/qcat?J/A+A/.}}
\label{sample}
\end{figure}

Since our 3.1\,mm data are of lower spatial resolution, we can smooth
the higher frequency data to this resolution ($1.9'' \times 1.4''$)
and investigate the SEDs of the relatively larger sub-components.
Figure \ref{sample} presents the compilation of all four continuum
images at the spatial resolution of the 3.1\,mm continuum dataset. We
clearly identify three (sub)mm continuum peaks at all wavelengths, the
formerly known sources mm1 to mm3. The vertical lines in Figure
\ref{sample} locate the R.A.~positions in the three lower-frequency
bands (3.1\,mm, 1.2\,mm and 875\,$\mu$m). While the positions of mm1
and mm3 coincide very well between the different bands, the position
of mm2 shifts progressively toward the south-east going from 3.1\,mm
to 875\,$\mu$m.  However, comparing the 1.2\,mm and 875\,$\mu$m
lower-resolution images in Figure \ref{sample} with the
higher-spatial-resolution images presented in Figure
\ref{1.3mm_850mu}, we see that the source mm2 consists of at least two
sub-sources, mm2a and mm2b. While the 1.2\,mm and 875\,$\mu$m
positions of mm2a coincide well (Fig.~\ref{sample} and full circles in
Fig.~\ref{sample}), the peak position of mm2b is shifted toward the
south west going from 1.2\,mm to the 875\,$\mu$m data. Based on the
lower-resolution 3.1\,mm image, it appears likely that at higher
spatial resolution mm2b would be even closer to mm2a going to even
longer wavelengths.  This positional shift of mm2b as well as the
non-detection of mm2c in the 875\,$\mu$m indicates that the dust
emission at these positions is not of typical protostellar nature, but
caused by other processes, e.g., the protostellar outflows in the
region. As already mentioned in \S \ref{small-scale}, the 3.1\,mm
image exhibits a smaller peak between mm1 and mm2 which is
approximately associated with mm2c in the shorter-wavelength
higher-resolution data (Fig.~\ref{1.3mm_850mu}). Taking additionally
into account the low spectral index toward this position (see below),
this feature is likely produced by an underlying jet/outflow
\citep{reynolds1986}, hence we labeled it the jet-feature in the
3.1\,mm image.

The 438\,$\mu$m image requires some additional explanations because
interferometric imaging at these frequency is still exploratory and
very difficult. The sub-source mm1 is at the phase center and the most
reliable feature. The emission peak of mm2 does not shift further to
the south-east as maybe expected from the lower-frequency data, but
the R.A. position coincides approximately with the 1.2\,mm peak
position whereas the Declination position is even closer to the
3.1\,mm position. Furthermore, the emission around mm2 in this band
extends a little bit further to the north than in the other images. We
are not entirely certain whether this is an artifact of the comparably
lower data quality in this band, or whether it may even be real
because many of the molecular line maps (e.g., CH$_3$OH, H$_2$CS or
C$^{34}$S) have an additional emission peak north of mm2 as well
(Leurini et al.~in prep.). Since high-spatial-resolution images at
these wavelengths are extremely rare so far (only 2 are published to
our knowledge, Orion-KL by \citealt{beuther2006a} and TW Hydra by
\citealt{qi2006}), it is hard to set additional constraints here.  The
location of mm3 is right at the edge of the primary beam of our
dataset (the phase center was chosen like that because we were also
interested in the CO(6--5) emission from the main outflow emanating
from mm1), and the measured fluxes toward mm3 are barely above the
noise ($\sim$2.5$\sigma$). To at least derive an upper limit for the
emission from mm3, we corrected the image shown in Figure \ref{sample}
for the primary beam attenuation. The derived fluxes shown in Table
\ref{masses} and used below for the mass/column density estimates, and
spectral energy distributions are measured from the
primary-beam-corrected image.

\subsection{Mass and column density estimates}

Assuming that the (sub)mm continuum emission is optically thin and
thus tracing the whole dust column densities with a gas to dust mass
ratio of 100, we can derive the gas column densities $N_{\rm{H_2}}$
and gas masses $M$ following \citet{hildebrand1983} and
\citet{beuther2002a}. The following sections will discuss the spectral
indices toward the various sub-regions in IRAS\,05358+3543 in more
detail, but for the mass calculations we adopt a dust opacity index
$\beta$ of 1.5, corresponding to dust opacities $\kappa$ of 0.2, 0.9,
1.4, and 4.0\,cm$^2$g$^{-1}$ at 3100, 1200, 875, and 438\,$\mu$m,
respectively.  Although the region contains gas components at various
temperatures (see Leurini et al.~in prep.), we assume an average
temperature of 50\,K for the mass estimates. Only for mm3, we use a
lower temperature of 20\,K as derived from the spectral energy
distribution shown in \S\ref{spectral_distribution}. Given the
  uncertainties of the dust opacities and the temperatures, we
  estimate the masses and column densities to be accurate within a
  factor 4. Table \ref{masses} presents the derived masses for the
various sub-sources estimated from the multi-wavelength dataset.

\begin{table}[htp]
\caption{(Sub)mm continuum fluxes and derived masses/column densities}
\begin{tabular}{lrrrr}
  \hline 
  \hline
  Source & $S_{\rm{int.}}$ & $S_{\rm{Peak}}$ & $M^e$ & $N_{\rm{H_2}}^e$ \\
  & [mJy] & $[\frac{\rm{mJy}}{\rm{beam}}]$ & [M$_{\odot}$] & [cm$^{-2}$] \\
  \hline
\multicolumn{5}{c}{3.1\,mm (AB) with $1.9''\times 1.4''$ beam}\\
\hline
mm1  & 20 & 10.3 & 10.6& 1.7\,$10^{24}$ \\
mm2  & 10 & 5.9  & 5.3 & 9.7\,$10^{23}$ \\
mm3$^d$  & 13 & 6.5  & 18.5 & 2.9\,$10^{24}$ \\
\hline  
\multicolumn{5}{c}{1.2\,mm (BCD) with $1.9''\times 1.4''$ beam}\\
\hline
mm1  & 215& 91.8 & 5.0 & 6.6\,$10^{23}$ \\
mm2  & 64 & 38.3 & 1.5 & 2.7\,$10^{23}$ \\
mm3$^d$  & 39 & 37.6 & 2.7 & 8.1\,$10^{23}$ \\
\hline  
\multicolumn{5}{c}{875\,$\mu$m with $1.9''\times 1.4''$ beam}\\
\hline
mm1  & 529& 270.8& 3.8 & 6.0\,$10^{23}$ \\
mm2  & 199& 110.1& 1.4 & 2.4\,$10^{23}$ \\
mm3$^d$  & 165& 137.5& 3.9 & 1.0\,$10^{24}$ \\
\hline
\multicolumn{5}{c}{438\,$\mu$m with $1.9''\times 1.4''$ beam$^b$}\\
\hline
mm1    &  8630 & 3589& 6.5 & 8.4\,$10^{23}$ \\
mm2    & 11450 & 3749& 8.6 & 8.8\,$10^{23}$ \\
mm3$^{c,d}$&   941 &  941& 3.2 & 9.9\,$10^{23}$ \\
\hline
\multicolumn{5}{c}{1.2\,mm (AB) with $0.60''\times 0.44''$ beam}\\
\hline
mm1a & 42 & 29.8 & 1.0 & 2.2\,$10^{24}$ \\
mm1b & 26 & 22.3 & 0.6 & 1.6\,$10^{24}$ \\
mm2a & 31 & 18.3 & 0.7 & 1.3\,$10^{24}$ \\
mm3$^d$  & 27 & 22.8 & 1.9 & 4.9\,$10^{24}$ \\
\hline
\multicolumn{5}{c}{1.2\,mm (BCD) with $1.26''\times 0.84''$ beam}\\
\hline
mm1a & 67 & 58.9 & & 1.1\,$10^{24}$ \\
mm1b & 62 & 54.8 & & 9.8\,$10^{23}$ \\
mm2a & 67 & 39.9 & & 7.2\,$10^{23}$ \\
mm2b & 33 & 33.3 & & 6.0\,$10^{23}$ \\
mm2c & 65 & 32.1 & & 5.8\,$10^{23}$ \\
mm2d & 41 & 26.5 & & 4.8\,$10^{23}$ \\
mm3$^d$  & 103& 48.0 & & 2.6\,$10^{24}$ \\
\hline
\multicolumn{5}{c}{875\,$\mu$m with $1.14''\times 0.57''$ beam}\\
\hline
mm1a & 225& 164.3& & 1.5\,$10^{24}$ \\
mm1b & 124& 106.4& & 9.6\,$10^{23}$ \\
mm2a & 97 & 62.5 & & 5.7\,$10^{23}$ \\
mm2b & 69 & 58.6 & & 5.3\,$10^{23}$ \\
mm2c & ?$^a$ \\
mm2d & 68 & 46.2 & & 4.2\,$10^{23}$ \\
mm3$^d$  & 212& 116.1& & 3.5\,$10^{24}$ \\
\hline  
\end{tabular}
~\\
{\footnotesize 
  $^a$ No clear emission peak in that band.\\
  $^b$ Fluxes in this band are measured from the primary-beam corrected image.\\
  $^c$ The flux values are only 2.5$\sigma$.\\
  $^d$ A colder temperature of 20\,K was used for mm3 (\S\ref{spectral_distribution}).\\
  $^e$ The masses and column densities are estimated to be accurate within a factor 4.
}
\label{masses}
\end{table}

As expected, the derived masses are all below the previous single-dish 1.2\,mm
and lower-resolution 3\,mm PdBI mass estimates (\S \ref{intro}). Some part of
the discrepancy can be solved by the different assumption for the dust opacity
$\beta$ used in the previous studies ($\beta = 2$,
\citealt{beuther2002a,beuther2002d}) and here ($\beta = 1.5$). However, the
difference in $\beta$ accounts to a mass difference of approximately a factor 2
at 1.2\,mm and 875\,$\mu$m, and the larger effect is the missing flux in the
new interferometer data.  The integrated single-dish fluxes at 1.2\,mm and
850\,$\mu$m are 6\,Jy and 28.7\,Jy \citep{williams2004}, respectively, whereas
the integrated fluxes in our high-spatial-resolution 1.2\,mm (BCD
configuration) and 875\,$\mu$m images (Fig.~\ref{1.3mm_850mu}) are 0.85\,Jy
and 1.37\,Jy, respectively.  Thus, we are recovering only between 15\% and 5\%
of the flux - and hence mass - in these interferometric images.  The
difference in flux recovery can already be recognized by just comparing the
values for the 1.2\,mm data in the AB and BCD configurations, respectively
(Table \ref{masses}). Therefore, we omit the mass estimates of the
higher-resolution data in Table \ref{masses}. 

An additional effect arises from the various wavelength observations:
Comparing the three mass estimates derived from 3.1\,mm, 1.2\,mm and
875\,$\mu$m data with the same spatial resolution ($1.9''\times 1.4''$, Table
\ref{masses}), the masses at 1.2\,mm and 875\,$\mu$m are approximately the
same, whereas the masses obtained from the 3.1\,mm data are more than a factor
2 higher. We will discuss the spectral energy distributions in detail below,
but already this comparison indicates that at 3\,mm additional non-dust
contributions to the continuum flux have to be taken into account.

Regarding the flux measurements and mass/column density estimates in
the 438\,$\mu$m band, the values for mm1 are reasonable compared to
the other bands. Since the 438\,$\mu$m flux for mm3 is only
$2.5\sigma$ limit, the derived mass and column density can be
considered as an upper limit. The strangest values are for mm2 in the
438\,$\mu$m band because they are higher than for mm1 whereas this is
always the other way round in the longer wavelength bands. As
discussed in the previous section, with the current data is is
difficult to decide whether this is an artifact of the dataset, or
whether there are additional contributions to the continuum emission
in this band as maybe suggested by the strong line emission peaks a
small distance to the north of mm2 (Leurini et al.~in prep.).

\subsection{Spectral energy analysis from the 
multi-wavelength (sub)mm continuum images}
\label{spectral_distribution}

\begin{figure*}[htb]
\begin{center}
\includegraphics[angle=-90,width=6.85cm]{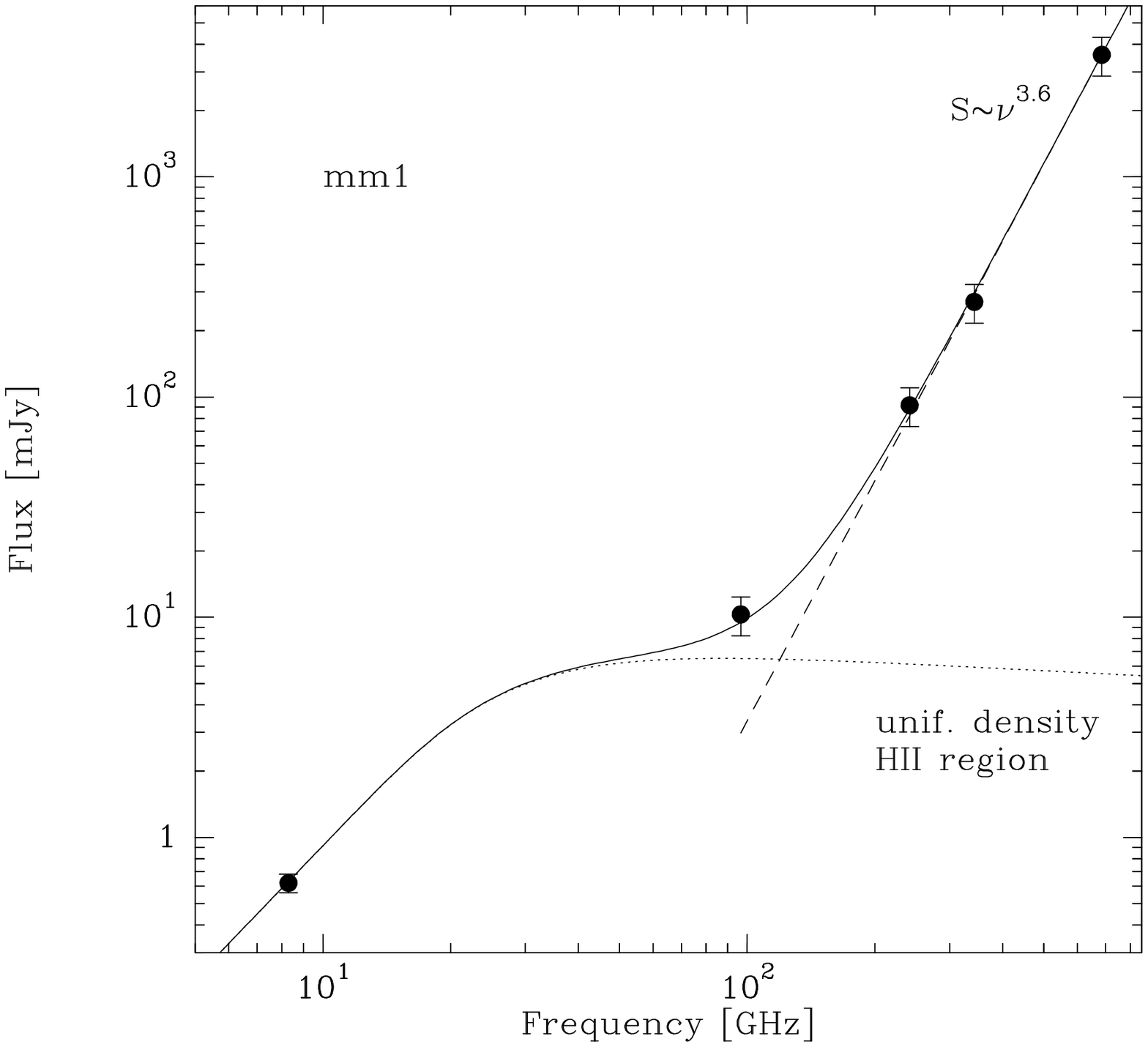}
\includegraphics[angle=-90,width=6.15cm]{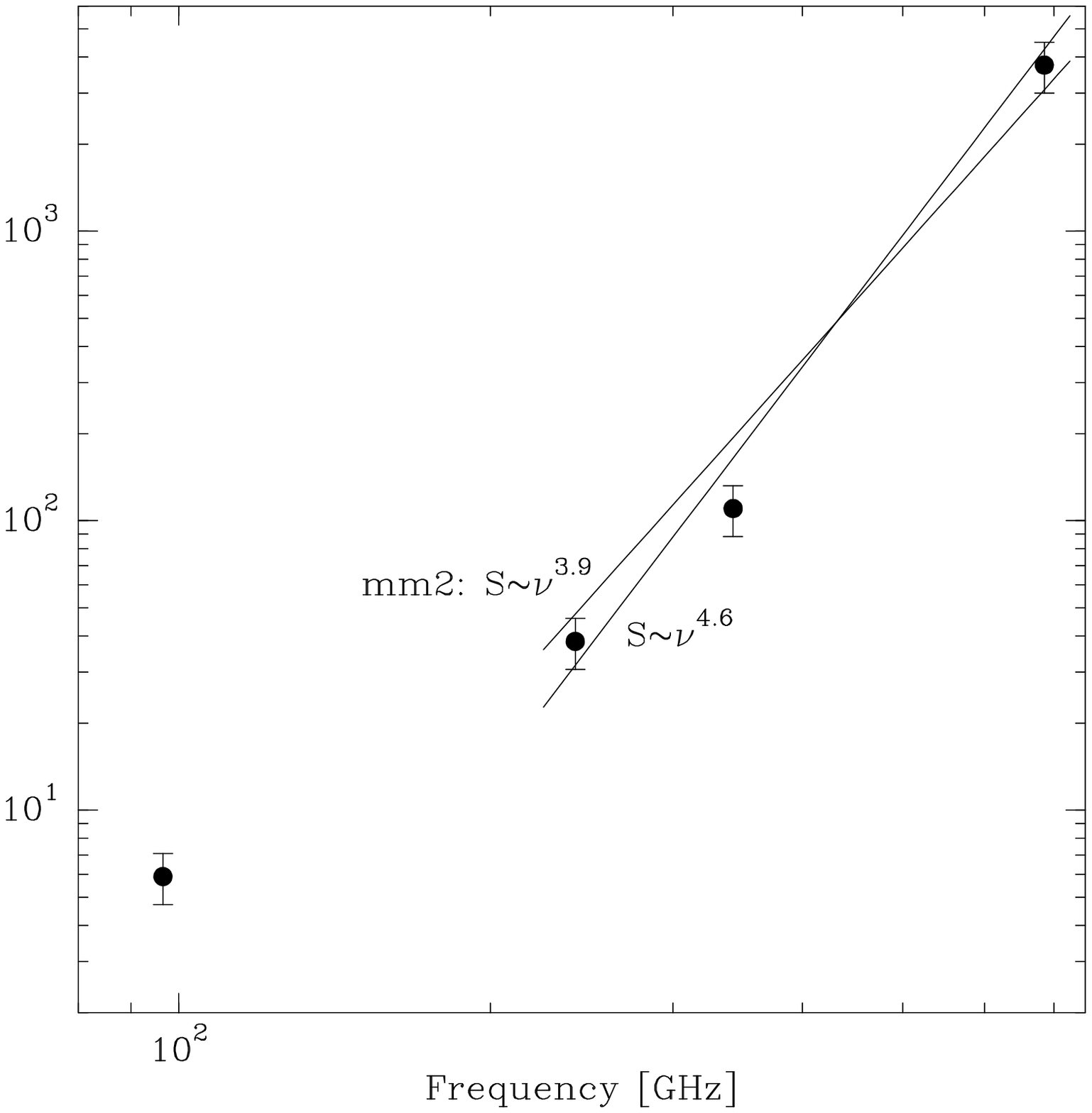}\\
\includegraphics[angle=-90,width=14cm]{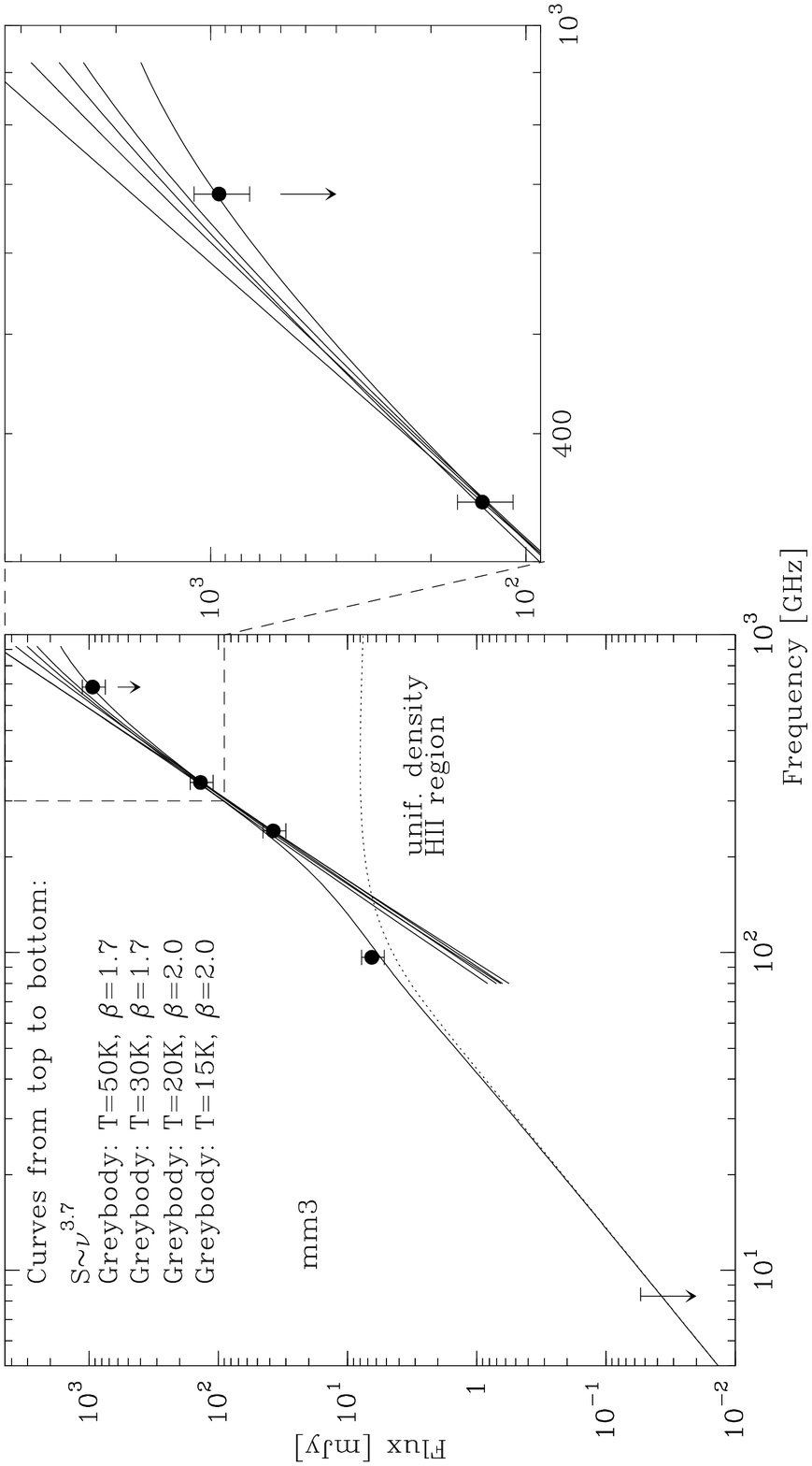}
\end{center}
\caption{Spectral energy distributions (SEDs) for mm1, mm2 and mm3. For mm1
  and mm3, the data are fitted with a uniform density H{\sc ii} region at low
  frequencies and a power-law with exponent 3.6 and 3.7 in the (sub)mm part of
  the spectrum, respectively.  For mm3, we present also various grey-body
  functions varying the temperature $T$ and the dust emissivity index $\beta$
  as labeled in the plot. The arrows in the mm3 SEDs mark the upper limits
  ($1\sigma$ noise at 8.3\,GHz, and the tentative 941\,mJy detection
  corresponding to $2.5\sigma$ at 685\,GHz). For mm2, we show two extreme
  power-laws fitting the 1.2\,mm and the 438\,$\mu$m but missing the
  865\,$\mu$m point.  We measured the fluxes toward the peak positions of each
  image in Fig.~\ref{sample}, respectively.}
\label{sed}
\end{figure*}

In spite of the complexity of the region, we can derive the SEDs from the
lower-resolution images toward mm1, mm2 and mm3 (Fig.~\ref{sample}). The
emission in the 3.1\,mm and 438\,$\mu$m bands appears a little more extended
which may be due to a better sampling of the uv-plane at the corresponding
scales.  Therefore, we refrain from using integrated fluxes for the various
positions but rather use the peak fluxes listed in Table \ref{masses}.  The
peak fluxes are derived at the peak position of each image. Figure \ref{sed}
presents the final (sub)mm SEDs for the three positions at the given spatial
resolution.

For all three (sub)mm peaks mm1 to mm3, the 3.1\,mm point is
consistently too high to be fitted with a single power law which is
expected for pure dust continuum emission at (sub)mm wavelength.
Therefore, the (sub)mm power-law fits at all three positions were
derived from the shorter wavelength data only.

{\it Source mm1:} The 1.2\,mm to 438\,$\mu$m data are well fitted by a
single power-law $S\propto \nu^{\alpha}$ with $\alpha\sim 3.6 \pm
0.3$.  This power-law exponent corresponds in the Rayleigh-Jeans
regime to a dust opacity index $\beta=\alpha - 2$ of $1.6\pm 0.3$,
observed often in high- and low-mass star-forming regions (e.g,
\citealt{andre2000,beuther2006b}).  The low-frequency 8.3\,GHz data
from the VLA are likely from a hypercompact H{\sc ii} region.
Therefore, we can fit the whole spectrum from 3.6\,cm to 438\,$\mu$m
with a uniform density hypercompact H{\sc ii} region at low to
intermediate frequencies and a dust component in the short mm to submm
range (Fig.~\ref{sed} top-left panel).  This is a typical spectrum one
expects from a young massive protostar where the central source has
already started hydrogen burning. Following \citet{rohlfs2006}, we can
calculate the emission measure (EM) using the approximate turnover
frequency of 35\,GHz and en electron temperature of $\sim 10^4$\,K.
The derived EM of $\sim 5.4\times 10^9$\,cm$^{-6}$ pc is very high,
about one to two orders of magnitude higher than values found in
typical ultracompact H{\sc ii} regions (e.g., \citealt{kurtz1994}).
Taking the optical thin free-free flux of $\sim$6.4\,mJy at 60\,GHz
(Fig.~\ref{sed}) and following \citet{kurtz1994}, we calculate a Lyman
continuum flux of $\sim 10^{45.4}$\,photons\,s$^{-1}$, which
corresponds approximately to a B1 Zero-Age-Main-Sequence star with a
luminosity of $\sim 10^{3.72}$\,L$_{\odot}$ and a stellar mass of
$\sim 13$\,M$_{\odot}$ \citep{panagia1973,lang1992}. Based on the
strong, collimated outflow that emanates from the vicinity of mm1 with
and outflow rate of $6\times 10^4$\,M$_{\odot}$ and a dynamical
timescale of $3.6\times 10^4$\,yrs \citep{beuther2002d}, it is likely
that this source is still in its accretion process, and therefore, we
still call it a protostar. The luminosity derived from the free-free
emission is of the same order as the total luminosity ($\sim
10^{3.8}$\,L$_{\odot}$) derived from the IRAS data \citep{sridha},
indicating that the luminosity of the whole region is really dominated
by the most massive protostar.

{\it Source mm3:} We can fit the 1.2\,mm and 875\,$\mu$m data
reasonably well with a power-law with $\alpha\sim 3.7\pm 0.3$
(corresponding to $\beta\sim 1.7\pm 0.3$). The 3.1\,mm fluxes are too
high for that fit.  Taking our 8.3\,GHz VLA $3\sigma$ upper limits of
56\,$\mu$Jy, it is again possible to fit the low-frequency part of the
spectrum with a hypercompact H{\sc ii} (Fig.~\ref{sed} bottom-left
panel). A point of caution here is that the turnover frequency from
the optically thick to the optically thin regime is relatively high,
between 70 and 100\,GHz. One possible way to explain this is that the
region is so young that the forming HCH{\sc ii} region is still
quenched and thus optically thick up to relatively high frequencies
without having the time and power to heat the surrounding gas.  Of
particular interest is the high-frequency data-point at 438\,$\mu$m
which is below the expectation of the power-law fit from the 1.2\,mm
and 875\,$\mu$m.  Since we have no cm nor compact line detection
toward mm3 (see \citealt{beuther2002d} and Leurini et al. in prep.),
this sub-source is a potential candidate for a still very cold massive
core at even earlier evolutionary stages. For such a cold core, the
peak of the SED moves to longer wavelength and one does not expect a
normal Rayleigh-Jeans-like power-law at short submm wavelengths
anymore.  Therefore we tried to fit the high-frequency SED with
grey-body functions with varying temperature (blackbody functions
modified by $1-e^{\tau_{\nu}}$ taking into account the finite optical
depth $\tau_{\nu}$ of the dust at all frequencies). The two
bottom-panels in Fig.~\ref{sed} show such grey-body functions with
temperatures between 50 and 15\,K (we also varied $\beta$ between 1.7
and 2).  One has to keep in mind that mm3 is at the edge of the
primary beam, that the imaging is difficult and that the data are
primary-beam corrected.  Nevertheless, these grey-body functions
indicate that most of the envelope dust and gas is likely at $\leq
20$\,K.

{\it Source mm2:} As already mentioned in the previous sections, mm2
is the strangest of the three sources. We do not find any reasonable
fit for the 1.2\,mm to 438\,$\mu$m. The fits presented in
Fig.~\ref{sed} (top-right panel) show the extreme power-laws fitting
the 1.2\,mm and the 438\,$\mu$m data. The 875\,$\mu$m data point is
below these fits. As pointed out previously, it may be possible that
the unproportionally high 438\,$\mu$m flux density could be partly an
artifact of this dataset. However, there may also be additional
contributions in this band (e.g., from the adjacent molecular line
peak) which have to be taken into account separately. Furthermore, the
data used for these SEDs are of lower spatial resolution smearing out
mm2a and mm2b. Since mm2b may be caused by an outflow/jet with a lower
spectral index (see the following paragraphs, this will likely
increase the complication of interpreting mm2's SED (see also the
cases of L1157 or IRAS\,20293+3952, \citealt{gueth2003,beuther2004e}).
In principal, a flattening of the SED in the long wavelength regime
could also be produced by a change of the dust grain size distribution
towards larger fractions of large grains within a protostellar disk.
However, this appears an unlikely scenario here because a low spectral
index is also observed in the larger-scale envelope of the region
\citep{williams2004}. Based on these data and the complicated
structure of the region, we are not able to draw further conclusions
from the SED of mm2.

\begin{figure}[htb]
\includegraphics[angle=-90,width=8.8cm]{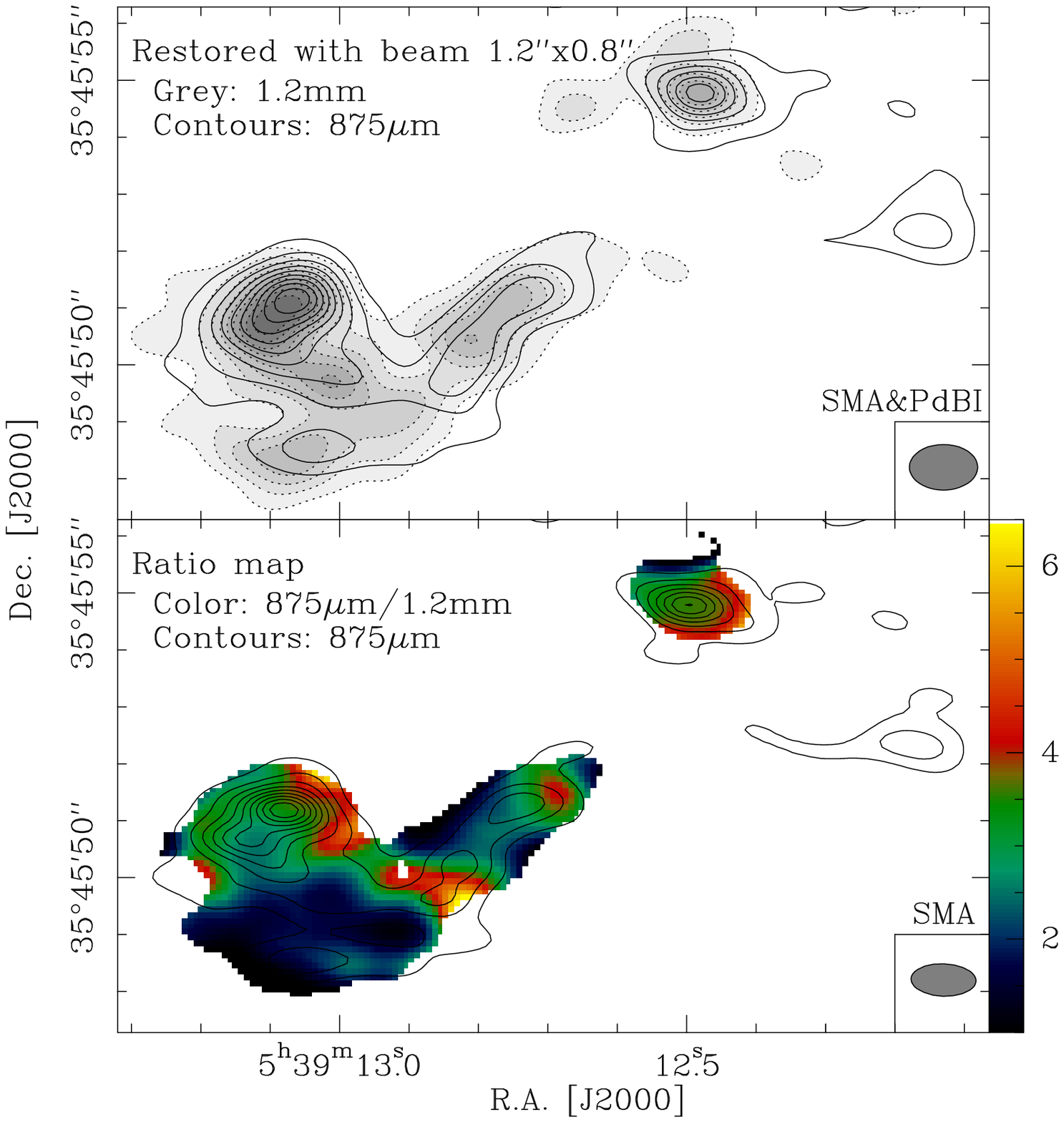}
\caption{Ratio map of the 875\,$\mu$m versus the 1.2\,mm data. Both
  datasets were smoothed to a common spatial resolution of
  $1.2''\times 0.8''$. To better align the two maps before producing
  the ratio, the 875\,$\mu$m was shifted by $0.18''$ to the west. The
  top panel shows the two maps at the same spatial resolution. The
  bottom panel then presents the ratio map in color scale with the
  875\,$\mu$m data at original resolution overlayed in contours. The
  synthesized beam of the data used for the ratio map is shown at the
  bottom-right of the top panel, and the bottom-right of the bottom
  panel shows the synthesized beam of the 875\,$\mu$m map at original
  resolution. {\it The files for this figure are also available in
    electronic form at the CDS via anonymous ftp to
    cdsarc.u-strasbg.fr (130.79.128.5) or via
    http://cdsweb.u-strasbg.fr/cgi-bin/qcat?J/A+A/.}}
\label{ratiomap}
\end{figure}

To derive additional small-scale information about the (sub)mm dust
continuum emission, we used the higher-spatial-resolution 1.2\,mm and
875\,$\mu$m data to produce a spectral index map of the (sub)mm
continuum emission. Therefore, we restored both datasets with the same
synthesized beam of $1.2''\times 0.8''$. To account for absolute
positional differences due to non-perfect phase calibration, we
shifted the 875\,$\mu$m image by $0.18''$ to the east aligning the
peak-positions of mm1a and mm3. Then the ratio map was constructed
from these two images (Figure \ref{ratiomap}). The spatial
distributions of the two restored maps shown in the top-panel of
Figure \ref{ratiomap} show that the spatial scales traced in both
bands are approximately the same, and thus a comparison of both images
appears reasonable.  Taking into account the flux uncertainties of
20\%, the different uv-sampling and the positional shift, we estimate
the uncertainties of this spectral index map of the order $\pm
0.5$. Edge effects of rising spectral index toward the north-east of
mm1a or toward the south-west of mm1b are most likely an artifact.
Similarly, the contrast of a low spectral index north of mm3 and a
high spectral index south of mm3 are potential artifacts caused by the
division of the different datasets. However, the variations of the
spectral indices throughout the region which are not edge effects are
likely to be real.

The spectral indices we derive from Figure \ref{ratiomap} toward the
four protostellar condensations mm1a, mm1b, mm2a, and mm3 (Table
\ref{positions}) are 3.6, 2.7, 2.3, and 3.6, respectively.  The
spectral indices toward mm1a and mm3 -- approximately the same as in
the lower-resolution data discussed before -- correspond to a rather
normal dust opacity index $\beta$ of 1.6 (e.g., \citealt{draine2006}),
whereas the spectral indices toward mm1b and mm2a are significantly
lower.  Since we aligned the two wavelength images simultaneously
with the peak positions from mm1a and mm3, the lower values toward the
other sources are likely to be real and no artifact from the shifting.
The positional shifting of mm2b from 1.2\,mm to 875\,$\mu$m
(Fig.~\ref{1.3mm_850mu}) is reflected in the ratio gradient toward
that region. Since mm2c is not detected at 875\,$\mu$m at all, the
corresponding spectral index is hence very low (of the order 1.5)
indicative of an underlying jet \citep{reynolds1986}.  Another
interesting position appears to be west of mm2a, where particularly
high values $> 4$ are found in the ratio map. Figure \ref{ratiomap}
clearly demonstrates the complexity of spectral index studies and the
large spread in spectral indices over very small spatial regions.
Possible physical implications will be discussed in \S \ref{indices}.

\subsection{Radial structure analysis from the
multi-wavelength (sub)mm continuum uv-data}
\label{uv}

\begin{figure}[htb]
\includegraphics[width=8.8cm]{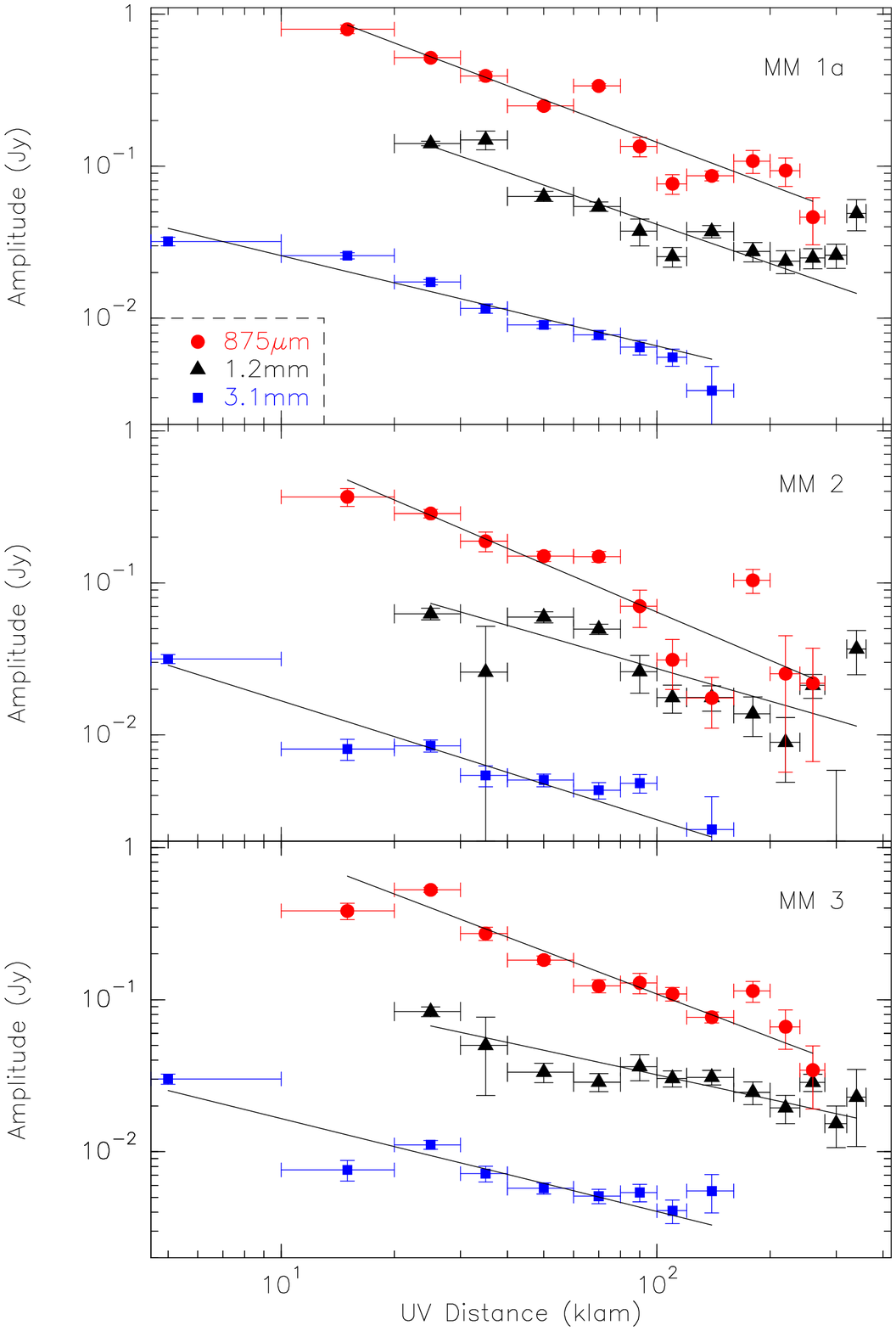}
\caption{Plots of amplitude versus uv-distance for the three main
  sub-sources in IRAS 05358+3543 at three observed frequencies.
  Results from power-law fits are shown as well. Although the top
  panel was produced centered on mm1a, the data $\leq$250\,k$\lambda$
  are from scales combining mm1a and mm1b. Therefore it resembles the
  intensity and density structure of this multiple system.}
\label{uvplots}
\end{figure}

To investigate the structure of the protostellar sub-sources detected
in the continuum, the data were analyzed directly in the uv-domain
avoiding the uv-sampling and cleaning difficulties.  To decrease the
influence of the other sub-sources on the analysis of a particular
object, two-dimensional Gaussians were fit as a first approximation to
the highest resolution images and the corresponding models were
removed from the uv-data, resulting in ``clean'' uv-datasets for each
of the three main sub-sources. Then the data for each sub-source were
binned in uv-annuli around the peak positions and averaged
vectorially.  The averaged amplitudes versus uv-radius are shown for
the 3 sources at the 3 observed wavelengths in Fig.~\ref{uvplots}.
Toward mm1, we selected the the mm1a peak position as the center, but
obviously most baselines trace structures combining mm1a and mm1b, and
only the largest baselines ($\geq$250\,k$\lambda$) are only due to
mm1a. Therefore, the derived intensity and density profile is for the
multiple system mm1a/mm1b. Most of the data can be fit reasonably well
by straight lines in the log-log plots, showing that the radial
structures of the sources are close to power-law profiles. The data
are inconsistent with Gaussians sources. Some of the data show
evidence for constant amplitudes at the largest uv-radii, hinting at
point source components at the smallest scales.  Deviations from the
straight lines are due to more complicated morphology imprinted on
power-law profiles, e.g. the substructures in mm1 and mm2.

Table~\ref{powerlaws} shows the results of the power-law fits to the
data. Assuming that the emission is due to dust at all frequencies,
the radial profiles should be the same at different frequencies.  The
differences that can be seen in the data therefore shows the
observational limitations of this method of radial profile analysis.
\citet{looney2003} give a relation between the power-law profiles in
the image and uv-planes, as well as how the observed uv-power-law
indices $V(s)$ are composed by the density and temperature structure:
$V(s) \sim s^{p+q-3}$, where $p$ and $q$ are the power-law indices of
the density $n \sim r^{-p}$ and temperature structure $T \sim r^{-q}$,
respectively. With a typical temperature power-law index of 0.4,
density indices from 1.5 to 2.0 translate to -1.1 to -0.6 in the uv
plane. Inspecting the observed values in Table~\ref{powerlaws} shows
that the millimeter cores have density profiles in this range, with
more observed indices closer to the $p=1.5$ than to the $p=2.0$
profile.

\begin{table}[htb]
\caption{Results of power-law fits to the continuum amplitudes versus
        uv-distances of the embedded protostellar sources}
\begin{center}
\begin{tabular}{lrrr}
\hline
\hline
 & mm1a$^a$ & mm2 & mm3  \\
\hline
$\alpha$(3.1mm) &  -0.59  &   -0.78  &   -0.61 \\
$\alpha$(1.2mm) &  -0.86  &   -0.71  &   -0.53 \\
$\alpha$(875$\mu$m) &  -0.93  &   -1.05  &   -0.94 \\
$\alpha$ av. & $-0.8\pm 0.2$ & $-0.85\pm 0.2$ & $-0.7\pm 0.2$ \\
$p$(3.1mm) &  2.0  &   1.8  &   2.0 \\
$p$(1.2mm) &  1.7  &   1.9  &   2.1 \\
$p$(875$\mu$m) &  1.7  &   1.5  &   1.7 \\
$p$ av.$^b$ & $1.8\pm 0.3$ & $1.75\pm 0.3$ & $1.9\pm 0.3$ \\
\hline
\hline
\end{tabular}
\end{center}
$^a$ The fit was centered on mm1a but it covers mainly structures combining mm1a and mm1b.\\
$^b$ The error for the average $p$ takes also into a account an error for $q$ of $\pm 0.2$.
\label{powerlaws}
\end{table}

\section{Discussion}
\label{discussion}

\subsection{Protostellar Powerhouses and other sources}
\label{power}

The double-structure mm1a and mm1b is separated by only $\sim$1700\,AU
projected on the plane of the sky. This double-system may well be a
massive bound binary system, similar to the the potential high-mass
protocluster or binary system W3(H$_2$O)
\citep{wyrowski1999,chen2006}. Considering also the third source
detected at mid-infrared wavelengths by \citet{longmore2006}, there
may be three sources within a projected circle with radius of
$\sim$1000\,AU. To carry the analogy with the W3 region even further,
we note that the ultracompact HII region/molecular hot core W3(OH)
lies at a similar distance from W3(H$_2$O) \citep{wyrowski1999} as
IRAS\,05358+3543 mm2 and mm3 from mm1a/b.  The measured expansion age
of the W3(OH) UCH{\sc ii} region is only 2300 years
\citep{kawamura1998}. It is well possible that IRAS\,05358+3543 which
now harbors a hypercompact H{\sc ii} region might evolve in to an
UCH{\sc ii} region on a similar time scale, whereas mm2 and mm3 might
evolve into regions resembling the present mm1a/b [and W3(H$_2$O)]
on similar time scales.

The two sources mm1a/b are located within one apparently common gas
and dust core, and the projected separation of $\sim$1700\,AU is
within the range of observed distances for proto-binary sources
\citep{launhardt2004}. From the molecular line data presented in
Leurini et al.~(in prep.) we are not able to derive a clear velocity
structure around the two cores which could be attributed to common
rotation (e.g., \citealt{launhardt2004}).  However, since we know
about the complexity of the region and two molecular outflows
emanating from the common mm1 core, it is not even a surprise that a
coherent velocity signature is missing. Each sub-source, mm1a and
mm1b, could drive one of two molecular outflows detected by
\citet{beuther2002d}.  Based on the higher flux of mm1a (Table
\ref{masses}), its association with the compact cm continuum source,
the Class {\sc ii} CH$_3$OH maser and the main mid-infrared source,
mm1a is the main power-house of the region likely driving the more
energetic large-scale collimated north-south outflow.

Interestingly, of the four mm2 sub-sources, only mm2a is detected as a
compact source in the highest resolution 1.2\,mm image. This is
indicative of mm2a being a genuine protostellar source whereas the
emission features associated with mm2b, mm2c and mm2d could be more
transient structures maybe caused by the underlying multiple outflow
system.

The source mm3 appears as the most simple object in the
IRAS\,05358+3543 massive star-forming region since it remains a single
peak and shows a spectral index indicative of protostellar dust
emission. The turnover of the SED at the shortest submm wavelengths
indicates that mm3 is the coldest sub-source in this region. Hence,
mm3 is a likely the youngest detected protostar in this system and may
potentially drive one of the multiple outflows identified in the
region.

While the CH$_3$OH Class {\sc ii} maser can clearly be associated with
mm1a, none of the four H$_2$O masers can be associated with any of the
protostellar sources. Since H$_2$O masers are believed to be produced
in shock fronts they do not necessarily have to be located in close
proximity to protostellar sources.

\subsection{Dust evolution and spectral indices}
\label{indices}

While dust opacity indices $\beta$ of 2 are usually found in the
interstellar medium, lower values are possible in star-forming cores.
For example, the dust opacity per unit dust mass $\kappa
(1.3\rm{mm})\sim 0.9$\,cm$^{2}$g$^{-1}$ suggested by
\citet{ossenkopf1994} for grains with thin ice mantles at gas
densities of $10^6$\,cm$^{-3}$ corresponds to $\beta \sim 1.4$.
Similarly, grain growth and dust coagulation in dense molecular cores
decreases the values for $\beta$ as well (e.g.,
\citealt{beckwith2000,draine2006}). A note of caution has to be added
that the effect of a low spectral index $\alpha$ can also be caused by
high optical depth from an underlying accretion disk and/or
hypercompact H{\sc ii} region, which can mimic low values of $\beta$.
Although it is difficult to differentiate between both scenarios, many
studies favor the grain-growth scenario.

The values $\beta$ of 1.6 and 1.7 measured toward mm1 and mm3 from the
lower-resolution dataset indicate that in these sub-sources the dust
has probably not evolved much yet and the dust properties are still
close to their properties in the interstellar medium. The SED of mm2
is too complex to derive a reasonable dust opacity index from the
given dataset.

The picture gets even more complicated going to the
higher-spatial-resolution ratio map presented in Figure
\ref{ratiomap}. Similar to the lower-resolution data, two of the
protostellar objects (mm1a and mm3) exhibit a rather normal spectral
index of 3.6, corresponding to a dust opacity index $\beta$ of 1.6.
While the spectral index of $2.7$ for mm1b is already lower, it still
is consistent with grain growth or high optical depth scenarios.

Although the very low value of $\alpha \sim 2.3$ toward the mm2a peak
would still be consistent with grain growth in circumstellar disks or
very high optical depths, the fact that we see a similarly low
spectral index on larger spatial scales \citep{williams2004} indicates
that additional physical processes may also come into play.  Since we
know of at least three molecular outflows emanating from the region,
these outflows/jets are the most likely culprits to change the
observed spectral indices. Thermal ionized wind models predict
spectral indices between 2 and $-0.1$ with typical spherical winds
exhibiting spectral indices $\alpha \sim 0.6$ \citep{reynolds1986}.
Since these winds turn optically thin going to higher frequencies,
their contribution to the SED is relatively larger at longer
wavelength. Therefore, adding a wind component on top of the expected
dust emission, a flattening of the spectral index is possible.
Spectral indices below 2 as observed toward mm2c and its surroundings
can be due solely to the outflows within the region (see also
\citealt{reynolds1986}).

Even more puzzling are the large spectral indices $> 4$ found in the
vicinity of mm2b and north-west of mm2a. Similarly high spectral
indices were observed toward the outflow region within the L1157
low-mass star-forming region by \citet{gueth2003} as well as toward
some parts of the Orion molecular ridge and a cold molecular cloud
near the Galactic Center \citep{lis1998,lis1998b}. While
\citet{gueth2003} attribute these high spectral indices to potentially
smaller dust grains and/or chemical properties (e.g.,
\citealt{kruegel1994,pollack1994}), \citet{lis1998} and
\citet{lis1998b} find that their high-$\beta$ locations are at low
temperatures. Therefore, they suggest that the high $\beta$ values can
be caused by dust grains covered with thick ice mantles (see also
\citealt{aannestad1975}). This picture is confirmed by recent modeling
of the dust opacity by \citet{boudet2005} indicating that $\beta$
decreases with increasing temperature, and that $\beta$ can be as high
as 2.5 (corresponding to a spectral index $\alpha =4.5$) at
temperatures of 15\,K. However, it is not obvious that the
temperatures toward the high-alpha positions in Fig.~\ref{ratiomap}
should be particularly low, and thus it is questionable whether this
picture can explain our observations. The fact that the positions of
mm2b do not coincide between the 1.2\,mm and the 875\,$\mu$m bands
also indicates that dynamic processes like protostellar outflows can
significantly alter the (sub)mm continuum emission.

It is also interesting to compare the spectral indices we get at high
spatial resolution with previous spectral index studies of the whole
region with single-dish SCUBA observations at the JCMT.
\citet{williams2004} derive a mean spectral index of 2 increasing
toward the (sub)mm peak to 2.6. This value is still significantly
lower than the 3.6 we measure toward mm1 which is the position of the
single-dish peak. Obviously, the single-dish data measure only
some kind of weighted average of the values one resolves with
interferometric observations.

\subsection{Density structure of massive star-forming cores}

The radial structure of the IRAS\,05358+3543 dust continuum on large
spatial scales probed by bolometer arrays has been studied by
\citet{beuther2002a} and \citet{williams2005}. These observations
resulted in radial density profiles of the cluster-forming gas and
dust clump with power-law distributions of $p$=1.9 and 1.5 for the
1.2\,mm and 850\,$\mu$m data, respectively. It is interesting that we
find approximately the same power-law distributions in the uv-analysis
of the individual sub-sources.  Hence, the density structure of the
large-scale cluster-forming clump appears to be similar to the
small-scale structure of its fragmented star-forming sub-cores.
Furthermore, the density distributions we find for the massive
star-forming cores resemble those derived previously for low-mass
star-forming regions (e.g., \citealt{motte2001}). The only other
high-mass study we know of that analyzed the structure of massive
star-forming cores with sub-arcsec resolution is by \citet{chen2006}
towards W3(H$_2$O) where they found density indices again close to
1.5. These results are consistent with the analytic models and
hydrodynamic simulations of massive cores by \citet{mckee2003} and
\citet{krumholz2006b}.

\subsection{Limitations and future prospects}

Although these observations are among the highest-spatial-resolution
multi-wavelength studies existing today, there are obviously still
limitations to what we can resolve. The finest spatial resolution of
$\sim 0.5''$ we achieve at 1.2\,mm corresponds at the given distance
of 1800\,pc to a linear resolution of 900\,AU, and we know from young
objects like W3IRS5 (e.g., \citealt{megeath2005}) to more evolved
regions like Orion (e.g., \citealt{preibisch1999}) that multiplicity
at scales $<1000$\,AU can well be present. \citet{koehler2006}
recently reported binary fractions for the Orion cluster of $23\pm
10\%$ for stars $>2$\,M$_{\odot}$ and $3.6\pm 3.2\%$ for stars
$<2$\,M$_{\odot}$. Even in IRAS\,05358+3543 we resolve mm1 into a
double-system (or triple system if including the MIR data) with
spatial separations of $\sim$1700\,AU (\S\ref{power}).  Therefore, all
parameters we derive for the various (sub)mm sub-source have to be
considered as potentially harboring multiple objects. This is
particularly important for the data at lower resolution ($>1''$),
where we know that they do not even resolve mm1a and mm1b properly.
For example, as mentioned in \S\ref{uv}, the radial density profile
derived toward the peak position of mm1a is actually the density
structure from the envelope encompassing mm1a and mm1b.  Similar
arguments hold for the SED analysis and the mass and column density
estimates.

Nevertheless, it is likely that the expected multiple unresolved
sub-sources are gravitationally bound and that they form from the same
gas and dust reservoir. Therefore, mass, column density, temperature
and density distribution of such entities are still the relevant
parameters for the physical properties of the collapsing cores. This
is consistent with recent 3D hydrodynamically simulations showing that
collapsing cores fragment less than previously expected, and that the
most massive fragments are likely forming within the massive accretion
disk \citep{krumholz2006b} which we do not resolve here anyway.

The only way to observationally assess the fragmentation of young and
cold massive star-forming cores in more detail is with higher spatial
resolution and higher sensitivity. While the spatial resolution of the
SMA and PdBI reaches today at best $\sim 0.3''$ (less than a factor 2
better than the presented observations), the brightness sensitivity
for these instruments at such scales is relatively poor. However, in
only a few years from now, the Atacama Large Millimeter Array (ALMA)
will be able to work regularly at $0.1''$ scales with good enough
sensitivity to detect even lower-mass members of the evolving
protoclusters. Then we will be able to derive the small
scale-fragmentation processes in better detail than possible today.

\section{Summary}

We have observed the very young massive star-forming region
IRAS\,05358+3543 interferometrically with high spatial resolution in
four frequency bands between 3.1\,mm and 438\,$\mu$m wavelength with
the Plateau de Bure Interferometer and the Submillimeter Array. The
observations reveal at least four sub-sources at the center of the
region that are likely of protostellar nature. One of them, mm1a, is
associated with CH$_3$OH Class {\sc ii} maser emission, a cm
wavelength hypercompact H{\sc ii} region and a mid-infrared source.
The center of two molecular outflows -- one of them the most
collimated massive molecular outflow observed so far in high-mass star
formation -- splits up into a double-source (mm1a and mm1b) with a
projected separation of 1700\,AU. With the given data, we cannot
distinguish whether this projected double-source is really a
proto-binary system or whether they are unbound. The additionally
detected mid-infrared source $\sim$1800\,AU south-west of mm1a
indicates a relatively high protostellar density with at least three
protostars in a projected circle with radius of $\sim$1000\,AU.  The
main power house of the region is the sub-source mm1a, but the other
sub-sources are likely to power outflows as well.  Furthermore, we
detect additional (sub)mm peaks (mm2b, mm2c and mm2d) which may not
necessarily be of protostellar nature but which could be caused by the
underlying outflows and jets. H$_2$O maser emission is found within
the region of dust continuum emission but no feature is clearly
associated with a protostellar (sub)mm continuum peak. This indicates
that H$_2$O maser emission is likely be produced in outflow shock
fronts.

This multi-frequency high-spatial-resolution (sub)mm continuum dataset
allows for the first time a spectral index analysis of individual
sub-sources in a massive evolving cluster that contain either
individual protostars or unresolved, likely gravitationally bound
multiple systems. The spectral energy distribution of the main power
house mm1 resembles a typical SED one expects for a massive
star-forming core: a hypercompact H{\sc ii} region from a
13\,M$_{\odot}$ B1 star is detected at cm wavelength whereas the mm to
submm spectrum is dominated by a cold dust component with a spectral
index of 3.6.  The 438\,$\mu$m data are of particular interest because
they fit well into the dust spectrum derived previously from the
longer wavelength data. The SED of the sub-source mm3 shows a
different behavior at this short wavelength: although we can only
derive upper limits at 438\,$\mu$m for mm3, this upper limit is
significantly below the power-law spectrum that one can fit from the
lower-frequency data and that would be expected for a normal
Rayleigh-Jeans-like part of the SED. Such a lower flux measurement can
be explained by even colder dust emission where the peak of the SED
shifts to longer wavelength and hence the 438\,$\mu$m point is already
closer to this peak and thus not on the Rayleigh-Jeans tail of the
Planck-function anymore.  Trying various grey-body functions, we
estimate an upper limit for the dust temperature of 20\,K.

A radial analysis of the data in the uv-domain reveals that the
density power-law indices $p$ vary for the different sub-sources
between 1.5 and 2, similar to observed density distributions of
low-mass star-forming cores. Furthermore, these power-law indices
resemble those previously found for the larger-scale cluster-forming
clump, indicative of similar density distributions on the high-mass
star- and cluster-forming scale.

Although the discussed sub-sources may still harbor multiple objects,
the derived physical parameters should represent the properties of
individual collapsing massive cores well.  Fragmentation on even
smaller spatial scales (e.g., within the massive accretion disks) will
likely only be observationally accessible with the advent of ALMA in a
few years from now.

\begin{acknowledgements} 
  We thank the PIs of the VLA cm-wavelength project AR482 Luis
  Rodriguez and Bo Reipurth for having no objections against us using
  these data. We are also thankful to Jan Martin Winters and Roberto
  Neri for their support during the reduction of the PdBI data.
  Furthermore, we acknowledge the referee's and the editor's reports
  that helped clarifying this paper. H.B.~acknowledges financial
  support by the Emmy-Noether-Program of the Deutsche
  Forschungsgemeinschaft (DFG, grant BE2578).  \end{acknowledgements}


\end{document}